\documentclass[pra,aps]{revtex4}

\begin{document}
\title{Spinor equation for the $W^{\pm}$ boson}
\author{Ruo Peng WANG\footnotemark
\footnotetext{email: rpwang@pku.edu.cn}}
\affiliation{ Physics
Department, Peking University, Beijing 100871, P.R.China }
\date{\today}

\begin{abstract}
I introduce spinor equations for the $W^{\pm}$ fields. The
properties of these spinor equations under space-time transformation
and under charge conjugation are studied. The expressions for
electric charge and current and densities of the $W^{\pm}$ fields
are obtained. Covariant quantization conditions are established, and
the vacuum energy for the $W^{\pm}$ fields is found to be zero.

\end{abstract}


\maketitle

\section{Introduction}

In the standard electro-weak theory, the $W^{\pm}$ bosons are
described by a 4-vector field that satisfies the Proca equation
\cite{proca1, proca2}. But however this 4-vector field can not be
regarded as the $W^{\pm}$ boson field, because the charge density of
the $W^{\pm}$ bosons can not be expressed as inner products of this
field with its adjoint field. Recently, I introduced a spinor field
theory for the photon \cite{wang}, in which the photon field is a
spinor, and satisfies a equation that is equivalent to Maxwell's
equations together with the relations between the 4-vector potential
and electric and magnetic fields, and the Lorentz gauge condition
for the 4-vector potential. In this paper, I modify the spinor
equation for the massless photon field to describe charged massive
spin 1 fields, which is the case of the $W^{\pm}$ boson fields. The
properties of spinor equations for the $W^{\pm}$ fields under
space-time transformation and under charge conjugation are studied
in detail. Expressions for electric current density, momentum and
angular momentum of the $W^{\pm}$ fields are obtained. Covariant
quantization conditions for the $W^{\pm}$ fields are established,
and the vacuum energy for the $W^{\pm}$ fields is found to be zero.

The spinor equations for the $W^{\pm}$ fields are introduced in
Sec.\ref{sec:equ}, the Lagrangian density and expressions for
electric charge and current densities are presented in
Sec.\ref{sec:lag}. The quantization of the $W^{\pm}$ fields is
carried out in Sec.\ref{sec:qua}.

\section{The equation for the $W^{\pm}$ boson}
\label{sec:equ}

The equation for the $W^{\pm}$ boson can be written as:
\begin{equation}\label{equ_w}
    i \hbar \frac{\partial}{\partial x_0} \psi_{w^\pm}(x)
    = -i\hbar \vec \alpha_w \cdot \nabla \psi_{w^\pm}(x)
    \pm m_w c \beta_w \psi_{w^\pm}(x)
\end{equation}
where
\begin{equation}
    \psi_{w^\pm}=\left(\psi_{w^\pm1}\ \ \psi_{w^\pm2}\ \ \psi_{w^\pm3}\ \ 0\ \
        \psi_{w^\pm5}\ \ \psi_{w^\pm6}\ \ \psi_{w^\pm7}\ \ \psi_{w^\pm8} \ \
        \psi_{w^\pm9}\ \ \psi_{w^\pm10}\ \ \psi_{w^\pm11}\ \ 0\ \
        0\ \ 0\ \ 0\ \ \psi_{w^\pm16}
        \right)^T
\end{equation}
and
\begin{equation}
    x_0=c t.
\end{equation}
$m_w$  in Eqs. (\ref{equ_w}) is the rest mass of $W^\pm$ boson. We
have the following expressions for the matrix $\vec \alpha_w$
\begin{equation}
    \vec \alpha_{w} = \left( \begin{array}{c c}
    \vec \alpha_e & 0 \\ 0 & -\vec \alpha_e
    \end{array} \right) ,
\end{equation}
where
\begin{equation}
    \alpha_{e1}=\left(\begin{array}{cccc}
        0 & 0 &  0 & -i\sigma_2 \\
        0 & 0 & -i\sigma_2 & 0 \\
        0 & i\sigma_2 &  0 & 0 \\
          i\sigma_2 & 0 &  0 & 0
        \end{array}\right)\;,\;
    \alpha_{e2}=\left(\begin{array}{cccc}
        0 & 0  & 0 & -I_2 \\
        0 & 0  & I_2 & 0 \\
          0 & I_2  & 0 & 0 \\
          -I_2 & 0 & 0 & 0
        \end{array}\right)\;,\;
    \alpha_{e3}=\left(\begin{array}{cccc}
        0 & 0 &  i\sigma_2 & 0 \\
        0 &  0 &  0 & -i\sigma_2 \\
          -i\sigma_2 &  0 &  0 & 0 \\
            0 & i\sigma_2 & 0 & 0
        \end{array}\right)\;
\end{equation}
with
\begin{equation}
    \sigma_2=\left(\begin{array}{cc}
        0 & -i  \\
        i & 0
    \end{array}\right)\;\;
    \mbox{and}\;\;
    I_2=\left(\begin{array}{cc}
        1 & 0  \\
        0 & 1
    \end{array}\right).
\end{equation}
The following anti-commutation relations hold for the matrixes
$\alpha_{w1}, \alpha_{w2}, \alpha_{w3}$:
\begin{equation}
    \alpha_{wn} \cdot \alpha_{wm} + \alpha_{wn} \cdot \alpha_{wm}
    = 2 \delta_{nm} \;, \; n,m = 1, 2, 3.
\end{equation}
The matrix $\beta_{w}$ is define as:
\begin{equation}
    \beta_{w} = \left( \begin{array}{c c}
    0 & I_8 \\ I_8 & 0
    \end{array} \right) ,
\end{equation}
where $I_8$ is the $8 \times 8$ unit matrix. The matrix $\beta_{w}$
satisfies the following relations:
\begin{equation}\label{t_a}
    \beta_w \vec \alpha_w + \vec \alpha_w \beta_w = 0 \;\; \mbox{and}\;\;
    \beta_w^2 = 1.
\end{equation}
Therefore
\begin{equation}\label{rel}
    (-i\hbar \vec \alpha_w \cdot \nabla
    + m_w c \beta_w)^2 = -\hbar^2 \triangle + m_w^2 c^2,
\end{equation}
where
\begin{equation}
    \triangle \equiv \frac{\partial ^2}{\partial x_1^2}
    +\frac{\partial ^2}{\partial x_2^2}
    +\frac{\partial ^2}{\partial x_3^2}.
\end{equation}

It is easy to verify that  $\psi_{w^+}(x)$ and $\psi^*_{w^-}(x)$
satisfy the same equation. We request that both $\psi_{w^+}(x)$ and
$\psi_{w^-}(x)$ contain only positive frequency components.

The equation (\ref{equ_w}) is invariant under continuous space-time
transformations (See the Appendices). We have
\begin{equation}\label{lrntz}
    \psi^\prime_{w^\pm}(x^\prime) = \exp (\vec \varphi \cdot \vec \Lambda)
    \psi_{w^\pm}(x^\prime)
\end{equation}
with
\begin{equation}\label{lmbd}
    \vec \Lambda =
    \left(\begin{array}{cc} - \vec l & 0 \\
    0 &  \vec \alpha_e - \vec l
    \end{array}\right)
\end{equation}
and
\begin{equation}
    \vec \varphi = \frac{\vec v}{v}
    \Bigl( \ln \sqrt{1+\frac{v}{c}}-\ln \sqrt{1-\frac{v}{c}}\Bigr).
\end{equation}
for Lorentz transformations, and
\begin{equation}
    \psi_{w^\pm}^\prime(x^\prime) = \exp {(i \vec \phi \cdot \vec s_f)} \psi_{w^\pm}(x^\prime)
\end{equation}
with
\begin{equation}
    \vec s_f=\left(\begin{array}{cc}
        \vec s & 0      \\
        0      & \vec s
        \end{array}\right).
\end{equation}
under space rotations, where
\begin{equation}
    \vec s=\left(\begin{array}{cc}
        \vec \Sigma & 0      \\
        0      & \vec \Sigma
        \end{array}\right)\;,\;
    \vec l= \left(\begin{array}{cc}
        0 & i\vec \Sigma      \\
        -i\vec \Sigma & 0
        \end{array}\right),
\end{equation}
and
\begin{equation}
    \Sigma_{1}=\left(\begin{array}{cccc}
        0 & 0 &  0 & 0 \\
        0 & 0 & -i & 0 \\
        0 & i &  0 & 0 \\
            0 & 0 &  0 & 0
        \end{array}\right)\;,\;
    \Sigma_{2}=\left(\begin{array}{cccc}
        0 & 0  & i & 0 \\
        0 & 0  & 0 & 0 \\
           -i & 0  & 0 & 0 \\
            0 & 0  & 0 & 0
        \end{array}\right)\;,\;
    \Sigma_{3}=\left(\begin{array}{cccc}
        0 & -i &  0 & 0 \\
            i &  0 &  0 & 0 \\
            0 &  0 &  0 & 0 \\
            0 &  0 &  0 & 0
        \end{array}\right).
\end{equation}
The following commutation relation holds for $\vec s_{f}$:
\begin{equation}\label{scom}
    [s_{fn},s_{fm}] = i \sum_{p=1}^3 \varepsilon_{nmp}s_{fp} .
\end{equation}
Eq. (\ref{equ_w}) is invariant also under space inversion, time
reversal and charge conjugation. It is easy to verify that
$\tau_0\psi_{w^\pm}(x_0,-\vec x)$, $\tau_0\psi^*_{w^\pm}(- x_0,\vec
x)$ and $\beta_0\psi^*_{w^\pm}( x_0,\vec x)$ satisfy the same spinor
equation Eq. (\ref{equ_w}) as $\psi_{w^\pm}( x_0,\vec x)$. Where
\begin{equation}
    \tau_0 = \left( \begin{array}{c c}
    -\beta_e & 0\\ 0 & -\beta_e
    \end{array} \right),\;
    \beta_0 = \left( \begin{array}{c c}
    I_8 & 0\\ 0 & -I_8
    \end{array} \right),
\end{equation}
with the matrix $\beta_e$ given by
\begin{equation}\label{b_i}
    \beta_e = \left( \begin{array}{c c}
    I_4 & 0 \\ 0 & -I_4
    \end{array} \right),
\end{equation}
in which  $I_4$ is the $4 \times 4$ unit matrix.

The spinor equation (\ref{equ_w}) can also be rewritten as a sets of
vector and scalar equations. Let
\begin{equation}
    \vec E_{w^\pm} = \pm i \sqrt{\mu_w}(\psi_{w^\pm1}, \psi_{w^\pm2}, \psi_{w^\pm3}),\;
    \vec H_{w^\pm} = \pm i \sqrt{\mu_w}(\psi_{w^\pm5}, \psi_{w^\pm6}, \psi_{w^\pm7}),
\end{equation}
\begin{equation}
    \vec A_{w^{\pm}} = \frac{1}{\sqrt{\mu_w}}(\psi_{w^\pm 9}, \psi_{w^\pm 10}, \psi_{w^\pm 11}),
\end{equation}
and
\begin{equation}
    A_{w^{\pm}0} = \frac{1}{\sqrt{\mu_w}} \psi_{w^\pm16},\;
    S_{w^\pm}= \pm i \sqrt{\mu_w} \psi_{w^\pm8},
\end{equation}
then the equation (\ref{equ_w}) is equivalent to the following
groups of equations
\begin{eqnarray}\label{emw1}
    \frac{\partial }{\partial x_0} \vec E_{w^\pm}
    &=& \nabla \times \vec{H}_{w^\pm} + \nabla S_{w^\pm} + \mu_w^2 \vec A_{w^{\pm}} \\
    \frac{\partial}{\partial x_0} \vec{H}_{w^\pm}
    &=&-\nabla \times \vec{E}_{w^\pm}\\ \label{emw3}
    0&=&-\nabla \cdot \vec{H}_{w^\pm}  \\ \label{emw4}
    \frac{\partial }{\partial x_0} S_{w^\pm} &=& \nabla \cdot
       \vec{E}_{w^\pm} + \mu_w^2 A_{w^{\pm}0}
\end{eqnarray}
and
\begin{eqnarray}
      \frac{\partial}{\partial x_0}
    \vec A_{w^{\pm}} &=& - \nabla A_{w^{\pm}0}
    - \vec{E}_{w^\pm} \\ \label{pw1}
      0 &=&  \nabla \times \vec A_{w^{\pm}}
    - \vec{H}_{w^\pm}  \\  \label{pw2}
      \frac{\partial}{\partial x_0} A_{w^{\pm}0} &=&
     -\nabla \cdot \vec A_{w^{\pm}} - S_{w^\pm}, \label{pw3}
\end{eqnarray}
with
\begin{equation}
    \mu_w = m_w c \hbar^{-1}.
\end{equation}
By using the anticommutation properties of $\vec \alpha$ and
$\beta_w$, we have
\begin{equation}
    \left(\frac{\partial^2}{\partial x_0^2} - \Delta + \mu^2_{w}\right)
    \psi_{w^\pm}(x) = 0,
\end{equation}
which is equivalent to
\begin{equation}
    \left(\frac{\partial^2}{\partial x_0^2} - \Delta + \mu^2_{w}\right)
     \left[ \vec E_{w^\pm}, \vec H_{w^\pm},  S_{w^\pm}\right] = 0,
\end{equation}
and
\begin{equation} \label{proca}
    \left(\frac{\partial^2}{\partial x_0^2} - \Delta + \mu^2_{w}\right)
    \left[A_{w^{\pm}0},  \vec A_{w^{\pm}}
    \right]   = 0,
\end{equation}

The equation (\ref{proca}) is just the Proca equations for 4-vector
``potentials'' $A_{w^{\pm}} = ( A_{w^{\pm}0}, \vec A_{w^{\pm}})$.
One may observe that if we let $S_{w^\pm}=0$, then Eqs.
(\ref{emw1})-(\ref{pw3}) are reduced to the Maxwell-Proca equations.
But as the charged current of the weak interaction coupled with the
$W^{\pm}$ fields does not satisfy the condition of continuity,
$S_{w^\pm}$ can not be zero when the weak interaction is considered.

\section{Lagrangian density}
\label{sec:lag}

The equation for $W^\pm$ boson (\ref{equ_w}) can be derived from the
following Lagrangian density

\begin{equation}\label{lag}
   {\cal L}_{w^\pm} = \bar \psi_{w^\pm}
    \left[i \hbar \left(\frac{\partial}{\partial t}
    + c \vec {\alpha}_w \cdot \nabla \right) \mp m_w c^2 \beta_w \right]
    \psi_{w^\pm}
\end{equation}
where $\bar \psi_w(x) = \psi^\dag_w(x) \tau_1$ with
\begin{equation}
    \tau_1 = \left( \begin{array}{c c}
    0 & \beta_e \\ \beta_e & 0
    \end{array} \right),
\end{equation}
is the adjoint field. $I_8$ is the $8 \times 8$ unit matrix. The
Lagrangian density (\ref{lag}) is invariant under a global phase
change of the $W^\pm$ field $\psi_{w^\pm}(x)$ . This implies the
conservation of the electric charge of $W^\pm$ fields :
\begin{equation}
    Q_{w^\pm} = \pm e \int \rho_{w^\pm} d^3 \vec x,
\end{equation}
and
\begin{equation}
    \frac{\partial}{\partial t} \rho_{w^\pm} + \nabla \cdot \vec j_{w^\pm}=0,
\end{equation}
where
\begin{equation}\label{rho}
    \rho_{w^+}(x) =  e \bar \psi_{w^+} (x) \psi_{w^+}(x), \;
    \rho_{w^-}(x) = - e \bar \psi_{w^-} (x) \psi_{w^-}(x)
\end{equation}
are the charge densities of $W^\pm$ fields and
\begin{equation}
    \vec {j}_{w^+}(x) =  e c \bar \psi_{w^+}(x) \vec \alpha_w
    \psi_{w^+}(x),\;
    \vec {j}_{w^-}(x) = - e c \bar \psi_{w^-}(x) \vec \alpha_w
    \psi_{w^-}(x)
\end{equation}
are the current densities. $e>0$ is the negative value of the
electron's electric charge. Due to the fact that the $W^{+}$ boson
and the $W^{-}$ boson have opposite electric charge, and
$\psi_{w^+}(x)$ commutes with $\psi_{w^-}(x)$, the charge densities
and the current densities of $W^\pm$ fields do not contain cross
terms between $\psi_{w^+}(x)$ and $\psi_{w^-}(x)$. That means a
$W^{+}$ boson will not annihilate with a $W^{-}$ boson by
interacting with the photon field.

According to the relation between symmetries and conservation laws,
we may obtain the following expressions for the momentum $\vec P$
and the angular momentum $\vec M$ of the free $W^\pm$  field:
\begin{equation}
    \vec {P}_{w^\pm} = - i \hbar \int d^3 \vec{x} \bar \psi_{w^\pm}\nabla \psi_{w^\pm},
\end{equation}
and
\begin{equation}
    \vec {M}_{w^\pm} =  \int d^3 \vec{x} \bar \psi_{w^\pm} [\vec{x} \times
    (- i\hbar \nabla)] \psi_{w^\pm} + \int d^3 \vec{x} \bar \psi_{w^\pm}
    (\hbar \vec{s}_f) \psi_{w^\pm}.
\end{equation}
It is clear that $\vec s_f$ can be interpreted as the spin operator
of the $W^\pm$ fields.

The conjugate field of $\psi_{w^\pm}$ is
\begin{equation}
    \pi_{w^\pm} = \frac {\partial {\cal L}}{\partial \dot {\psi_z}}
    = i \hbar \bar \psi_{w^\pm}.
\end{equation}
The Hamiltonian of the $W^\pm$ fields now can be calculated:
\begin{eqnarray}\label{hml}
    {H}_{w^\pm} & = & \int d^3 \vec{x} \left(
     \pi_{w^\pm} \dot{\psi_{w^\pm}} - {\cal L}_{w^\pm} \right)
     \nonumber\\
    & = & \int d^3 \vec{x}  \bar \psi_{w^\pm} \left( -i\hbar c \vec {\alpha}_{w} \cdot \nabla
    \pm m_w c^2 \beta_{w}\right) \psi_{w^\pm}
\end{eqnarray}

\section{quantization of the $W^\pm$ field}
\label{sec:qua}

It is convenient to quantize the $W^\pm$ field in the momentum
space. To do this, we have to find firstly the plan wave solutions
of the  $W^\pm$ field. By substituting the following form of
solution
\begin{equation}\label{pln}
    \psi_{w^+}(x) \propto \exp{(-i k x)}\Psi^{+}{(\vec{k})}
\end{equation}
into the spinor equation (\ref{equ_w}), we find
\begin{equation}\label{eqw}
    \left( \vec{\alpha_w} \cdot \vec{k} - k_0 -
    \mu_w \beta_{w} \right)
    \Psi^{+}{(\vec{k})}= 0.
\end{equation}

Eq. (\ref{eqw}) permits three independent nontrial solutions with
$k_0 = \sqrt{| \vec{k} |^2 + \mu_w^2} $. They can be chosen as
\begin{eqnarray}
    \Psi^{+}_{0}(\vec k) &=&\frac{1}{\sqrt{2 \mu_w k_0}}
    \Bigl( \mu_w \hat k_1 \;\;  \mu_w \hat k_2 \;\; \mu_w \hat k_3 \;\; 0 \;\; 0
    \; \; 0 \;\; 0 \;\; 0 \;\;
    k_0 \hat k_1 \;\; k_0 \hat k_2 \;\; k_0 \hat k_3
    \;\;0 \;\; 0 \;\; 0 \;\; 0 \;\;
    |\vec k|  \Bigr)^T,
\end{eqnarray}
and
\begin{eqnarray}
    \Psi^{+}_{\pm 1}(\vec k) &=& \frac{1}{2\sqrt{\mu_w k_0}}
    \Bigl( k_0( q_1 \pm i r_1) \;\; k_0(q_2 \pm i r_2) \;\;
    k_0(q _3 \pm i r_3) \;\; 0 \;\; |\vec k| ( r_1 \mp i q_1) \; \;
    |\vec k| (r_2 \mp i q_2) \;\;
    \nonumber \\&&
    |\vec k| (r_3 \mp iq_3) \;\; 0 \;\;
    \mu_w(q_1 \pm i r_1) \;\; \mu_w(q_2 \pm i r_2) \;\;
    \mu_w(q_3 \pm i r_3) \;\;0 \;\; 0 \;\; 0 \;\; 0 \;\; 0 \Bigr)^T,
\end{eqnarray}
where $\hat k = \vec k/|\vec k|$, and $\vec q$ and $\vec r$ are two
vectors of unity satisfying the following conditions:
\begin{equation}
    \hat k \times \vec q = \vec r,\;\;
    \hat k \times \vec r = -\vec q,\; \;
    \vec q \times \vec r =\hat k,\;\; \mbox {and}\;\;
    \vec r(-\hat k) = - \vec r(\hat k).
\end{equation}
Similarly, we have
\begin{equation}\label{plnm}
    \psi_{w^-}(x) \propto \exp{(-i k x)}\Psi^{-}_h{(\vec{k})},
\end{equation}
where $\Psi^{-}{(\vec{k})}$ satisfies the equation
\begin{equation}\label{eqwm}
    \left( \vec{\alpha_w} \cdot \vec{k} - k_0 +
    \mu_w \beta_{w} \right)
    \Psi^{-}_h{(\vec{k})}= 0,
\end{equation}
with
\begin{eqnarray}
    \Psi^{-}_{0}(\vec k) &=&\frac{1}{\sqrt{2 \mu_w k_0}}
    \Bigl( -\mu_w \hat k_1 \;\;  -\mu_w \hat k_2 \;\; -\mu_w \hat k_3 \;\; 0 \;\; 0
    \; \; 0 \;\; 0 \;\; 0 \;\;
    k_0 \hat k_1 \;\; k_0 \hat k_2 \;\; k_0 \hat k_3
    \;\;0 \;\; 0 \;\; 0 \;\; 0 \;\;
    |\vec k|  \Bigr)^T,
\end{eqnarray}
and
\begin{eqnarray}
    \Psi^{-}_{\pm 1}(\vec k) &=& \frac{1}{2\sqrt{\mu_w k_0}}
    \Bigl( k_0( q_1 \pm i r_1) \;\; k_0(q_2 \pm i r_2) \;\;
    k_0(q _3 \pm i r_3) \;\; 0 \;\; |\vec k| ( r_1 \mp i q_1) \; \;
    |\vec k| (r_2 \mp i q_2) \;\;
    \nonumber \\ &&
    |\vec k| (r_3 \mp iq_3) \;\; 0 \;\;
    -\mu_w(q_1 \pm i r_1) \;\; -\mu_w(q_2 \pm i r_2) \;\;
    -\mu_w(q_3 \pm i r_3) \;\;0 \;\; 0 \;\; 0 \;\; 0 \;\; 0 \Bigr)^T,
\end{eqnarray}
$ \Psi^{\pm}_{h}(\vec k) $ are orthogonal:
\begin{equation}
    \bar \Psi^{+}_h (\vec{k})\Psi^{+}_{h^\prime}(\vec{k}) = {\Psi^{+}_h}^\dag
    (\vec{k})\tau_1
    \Psi^{+}_{h^\prime}(\vec{k}) = \delta_{ h h^\prime}, \;
    \bar \Psi^{-}_h (\vec{k})\Psi^{-}_{h^\prime}(\vec{k}) = {\Psi^{-}_h}^\dag
    (\vec{k})\tau_1
    \Psi^{-}_{h^\prime}(\vec{k}) = \delta_{ h h^\prime}
    ,
\end{equation}
and
\begin{equation}
    \bar \Phi_h (\vec{k})\Psi_{h^\prime}(\vec{k}) = 0.
\end{equation}
We also have
\begin{equation}
    \bigl( \hat{k} \cdot \vec s_f \bigr) \Psi^{+}_{h}(\vec k)
    = h \Psi^{+}_{h}(\vec k), \;
    \bigl( \hat{k} \cdot \vec s_f \bigr) \Psi^{-}_{h}(\vec k)
    = h \Psi^{-}_{h}(\vec k) ,\;
    \mbox{ with } h=0,\pm1,
\end{equation}
and
\begin{equation}
    \bar \Psi^{+}_h (\vec{k})\vec s_f ^2 \Psi^{+}_{h}(\vec k)
    =
    \bar \Psi^{-}_h (\vec{k})\vec s_f ^2 \Psi^{-}_{h}(\vec k)
    =s(s+1)=2.
\end{equation}
That means the $W^\pm$ bosons are of spin s=1.

Having plan wave solutions the $W^\pm$ field, we may now expand
$\psi_{w^\pm}(x)$ in plane waves
\begin{equation}\label{exp}
    \psi_{w^+}(x)=\sum_{\vec k} \frac{1}{\sqrt {V}} e^{-ikx}
    \sum_{h}
    \Psi^{+}_{h}{(\vec{k})} a_h(\vec k),\;
    \psi_{w^-}(x)=\sum_{\vec k} \frac{1}{\sqrt {V}} e^{-ikx}
    \sum_{h}
    \Psi^{-}_{h}{(\vec{k})} b_h(\vec k)
\end{equation}
with $k_0 = \sqrt{|\vec k|^2 + \mu_w^2}$. According to relations
(\ref{lag}) and (\ref{exp}), the Lagrangian of the photon field can
expressed as a function of the variables $ q_{h\vec k} (t) $:
\begin{equation} \label{lgf}
    L(t,q) = \sum_{\vec k h} \Bigl( \hbar q^{+\dag}_{h\vec k}(t)
    (i\frac{\partial}{\partial t} - c k_0 ) q^+_{h\vec k} (t)
    + \hbar q^{-\dag}_{h\vec k}(t)
    (i\frac{\partial}{\partial t}  - c k_0 ) q^-_{h\vec k} (t) \Bigr),
\end{equation}
with
\begin{equation}
    q^+_{h\vec k}(t) = a_h(\vec k) \exp{(-i c k_0 t)},\;
    q^-_{h\vec k}(t) = b_h(\vec k) \exp{(-i c k_0 t)} .
\end{equation}
The conjugate momentum of $q^{\pm}_{h\vec k}(t)$ can be calculated,
and we have
\begin{equation}
    p^+_{h\vec k}(t) = \frac {\partial L}{\partial \dot q^+_{h\vec k}( t)}
    = i \hbar a_h^\dag(\vec k) \exp{(i ck_0 t)}
\end{equation}
and
\begin{equation}
    p^-_{h\vec k}(t) = \frac {\partial L}{\partial \dot q^-_{h\vec k}( t)}
    = i \hbar b^\dag_h(\vec k) \exp{(i ck_0 t)}.
\end{equation}
By applying the quantization condition $ [ q^\pm_{h\vec k} ,
p^\pm_{h^\prime \vec k^\prime} ] = i \hbar \delta_{hh^\prime}
\delta_{\vec k \vec k^\prime}$ we find the following commutation
relations
\begin{equation}\label{quant1}
    [a_h(\vec k), a^\dag_{h^\prime}(\vec k^\prime)]=
    \delta_{hh^\prime}\delta_{\vec k \vec k^\prime}.
\end{equation}
and
\begin{equation}\label{quant2}
    [b_h(\vec k), b^\dag_{h^\prime}(\vec k^\prime)]=
    \delta_{hh^\prime}\delta_{\vec k \vec k^\prime}.
\end{equation}

$ a_{h}(\vec k)$ and $ a^\dag _{h}(\vec k)$ are just the
annihilation operator and creation operator of $W^+$ bosons , and $
b_{h}(\vec k)$ and $ b^\dag _{h}(\vec k)$ are the same operators  of
$W^-$ bosons.  The Hamiltonian of the $W^\pm$ field can also be
calculated. We obtain
\begin{equation}
    H = \sum_{\vec k} \sum_{h} p_{h\vec k}\dot q_{h\vec k} - L
    = \sum_{\vec k h}
    c \hbar k_0 \Bigl(a^\dag_h(\vec k)a_h(\vec k) + b^\dag_h(\vec k)b_h(\vec k) \Bigr).
\end{equation}
We observe that the vacuum energy of the $W^\pm$  field is zero.

The commutation relations for the $W^\pm$   field can be written in
a covariant form. According to the commutation relations
(\ref{quant1}) and  (\ref{quant2}), and the expression (\ref{exp}),
we have
\begin{equation}\label{quantc}
    [ \psi_{w^\pm l}^\dag (x), \psi_{w^\pm m}(x^\prime)]= D_{lm}(x-x^\prime),
    \;\;    \text{with }\; l,m =1,2,\cdots,8 ,
\end{equation}
with the $8\times 8$ matrix $D(x)$ given by the following expression
\begin{equation}
    D(x) = \frac{1}{(2\pi)^3 \mu_w}\int_{k_0 > 0}
    d^4 k \delta (k^2 - \mu_w^2)\left[k_0 \vec k \cdot \vec l
    + (\vec k \cdot \vec l) (\vec k \cdot \vec l)
    + \mu_w^2 I_z \right]
    e^{-ikx},
\end{equation}
where
\begin{equation}
    I_w = \left( \begin{array}{c c}
    I_{43} & 0 \\ 0 & 0
    \end{array} \right)
    \; \mbox{ with } \;
    I_{43} = \left( \begin{array}{c c c c}
    1 & 0 & 0 & 0 \\ 0 & 1 & 0 & 0 \\
    0 & 0 & 1 & 0 \\ 0 & 0 & 0 & 0
    \end{array} \right)
    .
\end{equation}
The replacement
\begin{equation}
    \frac{1}{V}\sum_{\vec k} \longrightarrow
    \frac{1}{(2\pi)^3}\int d^3 \vec k
\end{equation}
was used in obtaining the relation (\ref{quantc}). Under Lorentz
transformations, $D(x)$ transforms to
\begin{equation}
    D^\prime (x^\prime) = \exp \bigl(-\vec \varphi \cdot \vec l \bigr)
    D(x^\prime) \exp \bigl(-\vec \varphi \cdot \vec l \bigr).
\end{equation}
One can verify with no difficulty that by using the expansions
(\ref{exp}), the commutation relations (\ref{quant1}) and
(\ref{quant2}) can be derived from commutation relation
(\ref{quantc}). Therefore, the commutation relations (\ref{quant1})
and (\ref{quant2}) and the the commutation relation (\ref{quantc})
are equivalent.

\section{Conclusion}

I introduced a spinor field theory for the $W^{\pm}$ fields. The
electric charge densities of the $W^{\pm}$ bosons can be expressed
as inner products of spinor $W^{\pm}$ fields with their adjoint
fields. Expressions for electric current density, momentum and
angular momentum of the $W^{\pm}$ fields are obtained. The
expressions for electric current densities of $W^\pm$ fields do not
contain cross terms between $\psi_{w^+}(x)$ and $\psi_{w^-}(x)$,
therefore a $W^{+}$ boson will not annihilate with a $W^{-}$ boson
by interacting with the photon field. Covariant quantization
conditions for the $W^{\pm}$ fields are established, and the vacuum
energy for the $W^{\pm}$ fields is found to be zero.

\appendix

\section{Invariance of the spinor equation for  $W^{\pm}$ field under Lorentz transformations}
To show the invariance of Eq. (\ref{equ_w}), it is convenient to
write the spinor field  $\psi_{w^\pm}(x)$ as
\begin{equation}
    \psi_{w^\pm}(x)=\left(\begin{array}{c}\psi_{u}(x) \\ \psi_{d}(x)
    \end{array} \right),
\end{equation}
where
\begin{equation}
    \psi_{u}=\left(\psi_{w^\pm1}\ \ \psi_{w^\pm2}\ \ \psi_{w^\pm3}\ \ 0\ \
        \psi_{w^\pm5}\ \ \psi_{w^\pm6}\ \ \psi_{w^\pm7}\ \ \psi_{w^\pm8}
        \right)^T, \;\;
    \psi_{d}=\left( \psi_{w^\pm9}\ \ \psi_{w^\pm10}\ \ \psi_{w^\pm11}\ \ 0\ \
        0\ \ 0\ \ 0\ \ \psi_{w^\pm16}
        \right)^T.
\end{equation}

According to Eq. (\ref{equ_w}), we have the following equations for
$\psi_{u}(x)$ and $\psi_{d}(x)$:
\begin{equation}\label{uw}
    \frac{\partial}{\partial x_0} \psi_{u}(x)=
    -\vec{\alpha}_e \cdot \nabla \psi_{u}(x) \mp i \mu_w
    \psi_{d}(x),
\end{equation}
and
\begin{equation}\label{dw}
    \frac{\partial}{\partial x_0} \psi_{d}(x)=
    \vec{\alpha}_e \cdot \nabla \psi_{d}(x) \mp i \mu_w
    \psi_{u}(x).
\end{equation}
The invariance of Eq. (\ref{equ_w}) is then equivalent to the
invariance of Eqs. (\ref{uw}) and (\ref{dw}).

By direct verification, one may find the following relations for
matrices $\vec s$ and $\vec l$:
\begin{equation}
    [s_{n},\alpha_{em}]= i \sum_{p=1}^3 \varepsilon_{nmp}
    \alpha_{ep}     \;,\;   n,m=1,2,3,
\end{equation}
and
\begin{equation} \label{lrel}
    \alpha_{en} l_m \alpha_{en} = l_m -(1-\delta_{nm})\alpha_{em}
    \;,\; n,m=1,2,3.
\end{equation}

Let's consider a Lorentz transformation
\begin{equation}
    \left(\begin{array}{c}x_1^\prime\\x_2^\prime
    \\x_3^\prime\\x_0^\prime \end{array} \right)
     = \left(\begin{array}{cccc}
    \cosh\varphi & 0 & 0 & - \sinh\varphi \\
    0 & 1 & 0 & 0 \\
    0 & 0 & 1 & 0 \\
    - \sinh\varphi & 0 & 0 & \cosh\varphi \end{array} \right)
    \left(\begin{array}{c}x_1\\x_2\\x_3\\x_0 \end{array}
    \right).
\end{equation}
We have
\begin{equation}\label{cotr}
    \frac{\partial}{\partial x_1} =
    \cosh\varphi \frac{\partial}{\partial x_1^\prime} - \sinh\varphi
    \frac{\partial}{\partial x_0^\prime} \;,\;
    \frac{\partial}{\partial x_0} =
    \cosh\varphi \frac{\partial}{\partial x_0^\prime} - \sinh\varphi
    \frac{\partial}{\partial x_1^\prime}.
\end{equation}
By using the relations (\ref{cotr}), Eqs. (\ref{uw}) and (\ref{dw})
can be written as
\begin{equation}
    (\cosh\varphi - \sinh\varphi \alpha_{e1})
    \frac{\partial} {\partial x_0^\prime }\psi_{u}(x^\prime)
    +\Bigl[(\cosh\varphi \alpha_{e1} - \sinh\varphi)\frac{\partial}
    {\partial x_1^\prime}
    +\alpha_{e2} \frac{\partial}{\partial x_2^\prime}
    +\alpha_{e3} \frac{\partial}{\partial x_3^\prime}\Bigr]
    \psi_{u}(x^\prime) \pm i \mu_w
    \psi_{d} (x^\prime) =0.
\end{equation}
and
\begin{equation}
    (\cosh\varphi + \sinh\varphi \alpha_{e1})
    \frac{\partial} {\partial x_0^\prime }\psi_{d}(x^\prime)
    -\Bigl[(\cosh\varphi \alpha_{e1} + \sinh\varphi)\frac{\partial}
    {\partial x_1^\prime}
    +\alpha_{e2} \frac{\partial}{\partial x_2^\prime}
    +\alpha_{e3} \frac{\partial}{\partial x_3^\prime}\Bigr]
    \psi_{d}(x^\prime) \pm i \mu_z
    \psi_{u} (x^\prime) =0.
\end{equation}
But
\begin{equation}
    l_1 \alpha_{e1} = \alpha_{e1} l_1,\;
    \exp (\pm\varphi \alpha_{e1}) = \cosh\varphi \pm \sinh\varphi \alpha_{e1}
\end{equation}
thus
\begin{eqnarray}\label{utr}
    \frac{\partial }{\partial x_0^\prime }\exp \bigl( -\varphi l_1 \bigr)
    \psi_{u}(x^\prime) &=&
    -\Bigl[\alpha_{e1}
    \frac{\partial}{\partial x_1^\prime}
    + \exp \Bigl(\varphi (\alpha_{e1} - l_1) \Bigr)
    \Bigl(  \alpha_{e2}\frac{\partial}{\partial x_2^\prime}
    +\alpha_{e3}\frac{\partial}{\partial x_3^\prime} \Bigr)
    \exp \bigl( \varphi l_1 \bigr) \Bigr]
        \nonumber \\
    &&
    \exp \bigl( -\varphi l_1 \bigr)
    \psi_{u}(x^\prime)
    \mp i\exp \Bigl(\varphi (\alpha_{e1} - l_1) \Bigr) \mu_w
    \psi_{d} (x^\prime),
\end{eqnarray}
and
\begin{eqnarray}\label{dtr}
    \frac{\partial }{\partial x_0^\prime }\exp \bigl( \varphi (\alpha_{e1} - l_1)  \bigr)
    \psi_{d}(x^\prime) &=&
    \Bigl[\alpha_{e1}
    \frac{\partial}{\partial x_1^\prime}
    + \exp \Bigl( - \varphi l_1 \Bigr)
    \Bigl(  \alpha_{e2}\frac{\partial}{\partial x_2^\prime}
    +\alpha_{e3}\frac{\partial}{\partial x_3^\prime} \Bigr)
    \exp \bigl( - \varphi (\alpha_{e1} - l_1)  \bigr) \Bigr]
        \nonumber \\
    &&
    \exp \bigl( \varphi (\alpha_{e1} - l_1)  \bigr)
    \psi_{d}(x^\prime)
    \mp i \exp \bigl(-\varphi l_1 \bigr) \mu_w
    \psi_{u} (x^\prime),
\end{eqnarray}

According to the relation (\ref{lrel}), we have
\begin{equation}
    \alpha_{em} l_1^n  = (l_1 - \alpha_{e1})^n\alpha_{em}
    \;,\; m=2,3,
\end{equation}
thus
\begin{equation}
    \alpha_{em} \exp \bigl( \varphi l_1 \bigr) =
    \exp \Bigl( \varphi (l_1 - \alpha_{e1}) \Bigr)\alpha_{em}
    \;,\;
    \exp \bigl( \varphi l_1 \bigr) \alpha_{em}  =
    \alpha_{em}  \exp \Bigl( \varphi (l_1 - \alpha_{e1}) \Bigr)
    \;,\; m=2,3.
\end{equation}
Equations (\ref{utr}) and (\ref{dtr}) become then
\begin{equation}\label{uw1}
    \frac{\partial}{\partial x^{\prime}_0}
    \psi^{\prime}_{u}(x^\prime)=
    -\vec{\alpha}_e \cdot \nabla \psi^{\prime}_{u}(x^\prime)
    \mp i \mu_w  \psi^{\prime}_{d}(x^{\prime}),
\end{equation}
and
\begin{equation}\label{dw1}
    \frac{\partial}{\partial x^{\prime}_0}
    \psi^{\prime}_{d}(x^\prime)=
    \vec{\alpha}_e \cdot \nabla \psi^{\prime}_{d}(x^\prime)
    \mp i \mu_w \psi^{\prime}_{u}(x^\prime),
\end{equation}
where
\begin{equation}
    \psi^{\prime}_{u}(x^\prime)
    = \exp \bigl( - \varphi l_1 \bigr) \psi_{u}(x^\prime)
    ,\;\;
    \psi^{\prime}_{d}(x^\prime)
    = \exp \bigl( \varphi ( \alpha_{e1} - l_1 ) \bigr)
    \psi_{d}(x^\prime).
\end{equation}
By using the expressions for $\vec \alpha_e$ and $\vec l$, one can
verify the following relations:
\begin{equation}
    \psi^{\prime}_{u4}(x^\prime) = \psi_{u4}(x) \equiv 0,\;\;
    \psi^{\prime}_{u8}(x^\prime) = \psi_{u8}(x),
\end{equation}
and
\begin{equation}
    \psi^{\prime}_{d4}(x^\prime) = \psi_{d4}(x) \equiv 0,\;\;
    \psi^{\prime}_{d5}(x^\prime) = \psi_{d5}(x) \equiv 0,\;\;
    \psi^{\prime}_{d6}(x^\prime) = \psi_{d6}(x) \equiv 0,\;\;
    \psi^{\prime}_{d7}(x^\prime) = \psi_{d7}(x) \equiv 0.
\end{equation}

The equations (\ref{uw1}) and (\ref{dw1}) in the new reference frame
has exactly the same form as Eqs. (\ref{uw}) and (\ref{dw}), the
spinor fields $\psi^{\prime}_{u}(x^\prime)$ and $\psi_{d}(x^\prime)$
in the new reference frame has exactly the same form as the spinor
fields $\psi_{u}(x)$ and $\psi_{d}(x)$. Therefore Eqs. (\ref{uw})
and (\ref{dw}) are invariant under Lorentz transformations.

\section{Invariance of the spinor equation for  $W^{\pm}$ fields under space rotation}

Let's consider an infinitesimal space rotation
\begin{equation}
    x_0^\prime=x_0,\;\;
    x_n^\prime = x_n - \sum_{m,p=1}^{3} \varepsilon_{nmp} \delta_m x_p
    \;\;n=1,2,3.
\end{equation}
We have
\begin{equation}
    \frac{\partial}{\partial x_n} =
    \frac{\partial}{\partial x_n^\prime} -
    \sum_{m,p=1}^{3} \varepsilon_{pmn} \delta_m
    \frac{\partial}{\partial x_p^\prime} \;,\;
    n=1,2,3.
\end{equation}
The equation (\ref{equ_w}) can be written as
\begin{equation}\label{equrot}
    i \hbar \frac{\partial}{\partial x_0^\prime}\psi_{w^\pm}(x^\prime) =
    -i \hbar \sum_{n=1}^{3} \Bigl[\alpha_{wn}
    - \sum_{l,m=1}^{3} \varepsilon_{nlm} \delta_l
    \alpha_{wm} \Bigr]
    \frac{\partial}{\partial x_n^\prime} \psi_{w^\pm}(x^\prime)
    \pm m_w c \beta_w \psi_{w^\pm}(x^\prime).
\end{equation}
According to expressions for $\vec \alpha_w$ and $\vec s_f$ we have
\begin{equation}
    \sum_{m=1}^3 \varepsilon_{nlm} \alpha_{wm} =
    i s_{fl}\alpha_{wn}- i \alpha_{wn} s_{fl},
\end{equation}
so
\begin{equation}
    \sum_{l,m=1}^3 \varepsilon_{nlm} \delta_l \alpha_{wm} =
    i \vec \delta \cdot \vec s_f \alpha_{wn} - i \alpha_{wn}
    \vec \delta \cdot \vec s_f .
\end{equation}
Therefore
\begin{equation}\label{equrot1}
    i\hbar \frac{\partial}{\partial x_0^\prime}
    (1 + i \vec \delta \cdot \vec s_f)\psi_{w^\pm}(x^\prime) =
    -i\hbar \sum_{n=1}^{3} \alpha_{en}
    \frac{\partial}{\partial x_n^\prime}
    (1 + i \vec \delta \cdot \vec s_f)\psi_{w^\pm}(x^\prime)
    \pm m_w c (1 + i \vec \delta \cdot \vec s_f) \beta_w \psi_{w^\pm}(x^\prime)
    .
\end{equation}
But $\beta_w \vec s_f = \vec s_f \beta_w$, so we may write Eq.
(\ref{equrot1}) as
\begin{equation}\label{equrot2}
    i\hbar \frac{\partial}{\partial x_0^\prime}
    \psi^{\prime}_{w^\pm}(x^\prime) =
    -i\hbar \sum_{n=1}^{3} \alpha_{en}
    \frac{\partial}{\partial x_n^\prime}
    \psi^{\prime}_{w^\pm}(x^\prime)
    \pm m_w c \beta_w \psi^{\prime}_{w^\pm}(x^\prime),
\end{equation}
with
\begin{equation}
    \psi^\prime_{w^\pm}(x^\prime) = (1 + i \vec \delta \cdot \vec s_f)
    \psi_{w^\pm}(x^\prime).
\end{equation}

The equation (\ref{equrot2}) has exactly the same form as Eq.
(\ref{equ_w}). So the spinor equation for $W^\pm$ field is invariant
under space rotations.

\end{document}